\documentclass[aps,twocolumn,notitlepage,superscriptaddress,nofootinbib]{revtex4-1}
\usepackage{amsmath}
\usepackage{epsfig}
\def\bea#1\eea{\begin{align}#1\end{align}} 
\newcommand{\nnu}{\nonumber\\}
\newcommand{\bwt}{\begin{widetext}}
\newcommand{\ewt}{\end{widetext}}

\begin{document}
\title{Next-to-leading order forward hadron production \\ in the small-$x$ regime: rapidity factorization}

\author{Zhong-Bo Kang}
\affiliation{Theoretical Division, 
                   Los Alamos National Laboratory, 
                   Los Alamos, NM 87545, USA}
                   
\author{Ivan Vitev}
\affiliation{Theoretical Division, 
                   Los Alamos National Laboratory, 
                   Los Alamos, NM 87545, USA}
                                      
\author{Hongxi Xing}
\affiliation{Theoretical Division, 
                   Los Alamos National Laboratory, 
                   Los Alamos, NM 87545, USA}
\affiliation{Institute of Particle Physics,
                   Central China Normal University, 
                   Wuhan 430079, China}

\begin{abstract}
Single inclusive hadron production at forward rapidity in high energy p+A collisions is an important probe of the high gluon density regime 
of QCD and the associated small-$x$ formalism. We revisit an earlier one-loop calculation 
to illustrate the significance of the
``rapidity factorization'' approach in this regime. Such  factorization separates the very small-$x$ unintegrated
gluon density evolution and leads to a new correction term to the physical cross section 
at one-loop level. Importantly, this rapidity factorization formalism remedies the previous unphysical negative next-to-leading order 
contribution to the cross section. It is much more stable with respect to ``rapidity'' variation 
when compared to  the leading-order calculation and provides improved 
agreement between theory and experiment in the forward rapidity region.
\end{abstract}

\date{\today}
\maketitle

%%%%%%%%%%%%%%
{\it Introduction.} As the theory of strong interactions, Quantum Chromodynamics (QCD)~\cite{Fritzsch:1973pi} has been extensively tested and 
verified. In particular, QCD in the weak coupling regime has been very successful in predicting and interpreting high energy scattering processes
in fixed target and collider experiments. Such a success is based on the well-established QCD collinear factorization formalism~\cite{Collins:1989gx}, 
which describes the hadron as a dilute system of partons. It was subsequently found that the parton densities (especially the gluon density) grow 
dramatically  when the longitudinal momentum fraction $x$ carried by a parton in a proton becomes very small due to bremsstrahlung processes. 
Such a fast growth would violate the fundamental principle of unitarity and cannot be sustained. It is, thus, expected that the gluon 
density will eventually become so large that a non-linear  regime, called a saturation regime~\cite{Gribov:1984tu}, will be reached. Another 
characteristic of the small-$x$ regime is that external hard probes  will interact with the partons in a nucleon or a nucleus coherently
rather than independently~\cite{Gelis:2010nm,Kang:2011bp}. In recent years, the high parton density limit has become one of the most  active research 
topics for QCD theory. The quest to identify the quantum coherent scattering regime is a critical goal for the ongoing experiments at 
the Relativistic Heavy Ion Collider (RHIC) and  the Large Hadron Collider (LHC). It is a corner stone of the physics program for  
the planned Electron Ion Collider (EIC)~\cite{Accardi:2012qut}.

Single forward hadron production in high energy proton-nucleus (p+A) collisions constitutes one of the key observables in searching for gluon 
saturation. The observed suppression of inclusive hadrons  at forward rapidity in d+Au collisions at RHIC~\cite{Arsene:2004ux} has provided 
evidence for the significance of cold nuclear matter effects, among them coherent multiple scattering. However, in the small-$x$  formalism,
experimental data are still mostly interpreted via leading order (LO) calculations~\cite{Dumitru:2002qt,Albacete:2010bs}. A significant step forward is the 
first calculation of forward hadron production at next-to-leading order (NLO)~\cite{Chirilli:2011km}. However, the resulting one-loop  
correction  in this approach is negative. At moderate and large transverse momenta it dominates the cross sections, which become 
negative (and unphysical)~\cite{Stasto:2013cha}.

In this paper we demonstrate that besides the well-known standard collinear factorization, which separates the short-distance dynamics from the 
long-distance physics, one has to pay close attention to the so-called ``rapidity factorization'' regime. It necessitates a rapidity cut-off 
to separate the very small-$x$ unintegrated gluon density evolution from the finite one-loop contributions. 
We revisit the NLO calculation for forward hadron production in high energy p+A collisions to show that such a  procedure leads to a new 
NLO correction term.  This term remedies the unphysical negative one-loop cross section obtained in~\cite{Stasto:2013cha}. The new 
formalism also leads to much less sensitivity to the choice of ``rapidity'' factorization scale at NLO in comparison to  LO  results and improved agreement 
between data and theory.

%%%%%%%%%%%%%%

{\it Rapidity factorization.} The mechanism of inclusive hadron production at forward rapidities in p+A collisions, $p+A\to h+X$, 
in the small-$x$ regime at LO can be described as follows: an energetic parton (either quark or gluon) from the proton scatters coherently 
on the gluon field of the nucleus, as it penetrates the target, and then fragments into the final-state hadron. Let us focus on the situation 
where a quark from the proton undergoes such scattering ($qA\to q$) to demonstrate the formalism. The 
differential cross section at forward rapidity $y$ and transverse momentum $p_\perp$ is given 
by \cite{Dumitru:2002qt,Albacete:2010bs}
\bea
\frac{d\sigma}{dyd^2p_\perp} & =  \int_{\tau}^1\frac{dz}{z^2}  D_{h/q}(z) x_p f_{q/p}(x_p) {\cal F}(x_g, k_\perp),
\label{LO}
\eea
where the sum over  quark flavors is suppressed for simplicity, $k_\perp = p_\perp/z$, and $\tau=\frac{p_\perp}{\sqrt{s}} e^y$. 
$f_{q/p}(x_p)$ is the collinear parton distribution function (PDF) in the proton with $x_p = \tau/z$ and $D_{h/q}(z)$ is the fragmentation function (FF). 
$x_g$ is the longitudinal momentum  fraction of the probed gluons in the nucleus and is given by $x_g=x_A$ with 
$x_A = \frac{p_\perp}{z\sqrt{s}} e^{-y}$. All the information for the transverse momentum transfer from  coherent 
multiple scattering is contained in ${\cal F}(x_g, k_\perp)$, the so-called unintegrated gluon distribution defined as
\bea
\hspace{-4pt} {\cal F}(x_g, k_\perp)=\int \frac{d^2 b_\perp d^2 b'_\perp}{(2\pi)^2} 
e^{-ik_\perp\cdot (b_\perp - b'_\perp)} S^{(2)}(b_\perp, b'_\perp),
\eea
where $S^{(2)}(b_\perp, b'_\perp)$ is the dipole scattering amplitude  given by $S^{(2)}(b_\perp, b'_\perp) 
= \frac{1}{N_c} \left\langle {\rm Tr}\left[U(b_\perp) U^\dagger(b'_\perp)\right]\right\rangle$. Here 
$U(b_\perp) = {\cal P} \exp\left\{ig_s \int_{-\infty}^{+\infty} dx^+ t^c A_c^-(x^+, b_\perp)\right\}$ is the Wilson line in the 
small-$x$ formalism. 

Let us now concentrate on the NLO calculation, in which we have to consider both real and virtual corrections. 
The calculation is standard in the so-called light-front perturbation theory~\cite{Bjorken:1970ah}, 
and the result can be written as the sum of three terms~\cite{Chirilli:2011km}, $d\sigma/dyd^2p_\perp=I^R+I^V+I^Y$. 
The expressions are given by
\bwt
\bea
I^R  =& \alpha_s C_F\int_{\tau}^1\frac{dz}{z^2}D_{h/q}(z)\int_{\tau/z}^1 d\xi \frac{1+\xi^2}{(1-\xi)_+} x f_{q/p}(x) 
\int\frac{d^2b_\perp d^2b'_\perp d^2x_\perp}{(2\pi)^4}e^{-ik_\perp \cdot (b_\perp-b'_\perp)}
\frac{2(x_\perp - b_\perp)\cdot(x_\perp - b'_\perp)}{(x_\perp - b_\perp)^2(x_\perp - b'_\perp)^2}
\nnu
&\times
\left[S^{(2)}(b_\perp, b'_\perp)+S^{(2)}(v_\perp, v'_\perp)-S^{(3)}(b_\perp, x_\perp, v'_\perp)-S^{(3)}(v_\perp, x_\perp, b'_\perp)\right],
\\
I^V = & -2 \alpha_s C_F \int_{\tau}^1\frac{dz}{z^2}D_{h/q}(z)x_p f_{q/p}(x_p) 
\int_{0}^1 d\xi \frac{1+\xi^2}{(1-\xi)_+}\int\frac{d^2v_\perp d^2v'_\perp d^2u_\perp}{(2\pi)^4}
e^{-ik_\perp\cdot (v_\perp - v'_\perp)}\frac{2}{u_\perp^2}
\nnu
&\times
\left[S^{(2)}(v_\perp,v'_\perp)-S^{(3)}(b_\perp, x_\perp, v'_\perp)\right],
\\
I^Y = &\int_{\tau}^1\frac{dz}{z^2} D_{h/q}(z) x_p f_{q/p}(x_p) 
\int\frac{d^2b_\perp d^2b'_\perp}{(2\pi)^2}e^{-i k_\perp\cdot (b_\perp - b'_\perp)} 
\nnu
&\times
\left\{\frac{\alpha_s N_c}{2\pi^2}  \int_0^1\frac{d\xi}{1-\xi}
\int d^2x_\perp \frac{(b_\perp - b'_\perp)^2}{(b_\perp - x_\perp)^2(x_\perp - b'_\perp)^2}
\left[S^{(4)}(b_\perp, x_\perp, b'_\perp) - S^{(2)}(b_\perp, b'_\perp)\right]\right\},
\label{Ycont}
\eea
\ewt
where $S^{(3)}$ and $S^{(4)}$ are multi-point gluon correlators defined in \cite{Chirilli:2011km,Marquet:2007vb}. 
Both $I^R$ and $I^V$ are finite when $\xi\to 1$ but they contain collinear divergences. To see the collinear divergences 
explicitly and, thus, regularize them, it is useful to work in  momentum space. We use dimensional 
regularization in $n=4-2\epsilon$ dimensions  with the change 
\bea
\int \frac{d^2 q_\perp}{(2\pi)^2} \to \mu^{2\epsilon} \int \frac{d^{2-2\epsilon} q_\perp}{(2\pi)^{2-2\epsilon}}.
\label{DR}
\eea
In momentum space the collinear divergences are manifest, as demonstrated in~\cite{Dumitru:2005gt,Chirilli:2011km}. They can be absorbed 
into the redefinition of either the PDF $f_{q/p}(x)$ or the FF $D_{h/q}(z)$, 
which leads to the well-known DGLAP evolution equations for PDFs and FFs \cite{Chirilli:2011km, Kang:2013raa}. 
Such a collinear factorization procedure introduces a factorization scale ($\mu$) dependence~\cite{scale}.  Since $\mu$ 
is an artificial scale, the physical cross section should not depend on it in an all-order result. 
In practice, since one can only calculate to finite order, some residual $\mu$-dependence remains. 
However, it should be reduced in the NLO calculation compared with the LO result and the cross section at NLO 
is expected to have smaller uncertainty \cite{Stasto:2013cha}. 

On the other hand, $I^Y$ is divergent in the limit $\xi\to 1$. This is the so-called rapidity divergence. 
It is instructive to realize that for forward hadron production we have 
\bea
\int_0^1\frac{d\xi}{1-\xi}=\int_0^1\frac{d\xi_g}{\xi_g}=  \int_0^{\infty}dy_g = \int_{-\infty}^{Y}dy_A,
\eea
where $\xi_g = 1-\xi$ is the momentum fraction of the projectile quark carried by the radiated gluon, with $y_g = \ln1/\xi_g$ the rapidity of the radiated gluon w.r.t. the projectile proton. On the other hand, $y_A = Y - y_g$ is the rapidity of the radiated gluon w.r.t. the target nucleus, where $Y = \ln (s/m_p^2)$ is the rapidity interval between the projectile proton and the target nucleus \footnote{Strictly speaking, $Y$ should be the rapidity interval between the projectile quark and the target nucleus. However, we are using the so-called hybrid formalism~\cite{Dumitru:2002qt,Albacete:2010bs,Chirilli:2011km}, in which the projectile quark is purely collinear to the parent proton without transverse momentum.  In this case we have quark momentum $p_q \approx x_p p$ with $p$ the proton momentum, and thus the quark rapidity is the same as the proton rapidity.}, with $s$ ($m_p$) the center-of-mass energy squared (nucleon mass). The divergence occurs when $y_A \to -\infty$, thus the name ``rapidity divergence''. Rapidity divergence is a general feature~\cite{Collins:2003fm} when one uses the transverse momentum dependent distributions, e.g.  ${\cal F}(x_g, k_\perp)$ in our case. It is very easy to see from Eq.~(\ref{Ycont})  that such a divergence disappears when one integrates over $k_\perp$ \cite{Chirilli:2011km}. Realizing that
\bea
\int_{-\infty}^{Y}dy_A = \int_{-\infty}^{Y_0} dy_A + \int_{Y_0}^{Y} dy_A,
\eea
following the ideas of collinear factorization, we compare $I^Y$ to the LO result in Eq.~\eqref{LO} and see 
that one should absorb this divergence into the redefinition of the dipole scattering amplitude 
\bea
S^{(2)}_{Y_0}(b_\perp, b'_\perp) &= S^{(2)}(b_\perp, b'_\perp) + \frac{\alpha_s N_c}{2\pi^2}\int_{-\infty}^{Y_0}dy_A
\nnu
&\times \int d^2x_\perp \frac{(b_\perp - b'_\perp)^2}{(b_\perp - x_\perp)^2(x_\perp - b_\perp')^2}
\nnu
&\times
\left[S^{(4)}(b_\perp, x_\perp, b'_\perp) - S^{(2)}(b_\perp, b'_\perp) \right].
\eea
Here, the rapidity cut-off $Y_0$ is introduced to separate the 
``fast'' and ``slow'' gluon fields~\cite{Balitsky:1998kc,Balitsky:1995ub}. The $Y_0$-dependence for the renormalized dipole 
gluon distribution $S^{(2)}_{Y_0}(b_\perp, b'_\perp)$ leads to the well-known Balitsky-Kovchegov (BK) evolution 
equation~\cite{Balitsky:1995ub,Kovchegov:1999yj}
\bea
\frac{\partial}{\partial Y_0}S^{(2)}_{Y_0}(b_\perp, b'_\perp)
&= \frac{\alpha_sN_c}{2\pi^2}\int d^2x_\perp \frac{(b_\perp - b'_\perp)^2}{(b_\perp - x_\perp)^2(x_\perp - b_\perp')^2}
\nnu
&\hspace{-20pt}  \times
\left[S^{(4)}_{Y_0}(b_\perp, x_\perp, b'_\perp) - S^{(2)}_{Y_0}(b_\perp, b'_\perp)\right].
\eea
After the subtraction, a finite correction appears from the rapidity factorization procedure
\bea
\Delta H_Y &=  \frac{\alpha_s N_c}{2\pi^2} \int_{\tau}^1\frac{dz}{z^2} D_{h/q}(z) x_p f_{q/p}(x_p)  \int_{Y_0}^{Y} dy_A 
\nnu
&\hspace{-12pt} \times
\bigg\{\int\frac{d^2b_\perp d^2b'_\perp  d^2x_\perp}{(2\pi)^2}
 \frac{(b_\perp - b'_\perp)^2}{(b_\perp - x_\perp)^2(x_\perp - b_\perp')^2}
\nnu
&\hspace{-12pt} \times e^{-i k_\perp\cdot (b_\perp - b'_\perp)} 
\left[S^{(4)}(b_\perp, x_\perp, b'_\perp) - S^{(2)}(b_\perp, b'_\perp) \right]\bigg\}.
\label{rap-corr}
\eea
As we will show later, it is this new correction term that was missed in~\cite{Stasto:2013cha} and which ensures     
that the NLO cross section is positive definite \footnote{In principle, one could choose the rapidity cut-off $Y_0=Y$ such
that the correction term $\Delta H_Y$ vanishes. However, in this case, one has to allow the dipole
gluon distribution $S^{(2)}(b_\perp, b_\perp')$ evolves up to rapidity $Y=\ln (s/m_p^2)$ instead to the 
typical gluon rapidity in the target nucleus $\sim \ln1/x_g$ as we will show below. Nevertheless, we could regard such
a choice as a different scheme~\cite{Balitsky:2012bs}.}. 
Similarly to the collinear factorization case, 
a rapidity cut-off scale $Y_0$ is introduced in  rapidity factorization. The physical cross section should 
also be independent of such a rapidity cut-off in the all-order result. In our finite order calculation 
some residual $Y_0$-dependence is expected to remain. However, it should be reduced at NLO when compared to the LO result.
One can  choose the gluon rapidity cut-off to be the one related to the gluon momentum fraction from the LO kinematics, e.g. $x_g=x_A$. 
Unlike the usual collinear factorization, which can be seen as separating  perturbative from nonperturbative physics, 
both rapidity separated parts have perturbative and nonperturbative contributions at the same time~\cite{Balitsky:1998kc}.

Let us now better understand  the rapidity correction term $\Delta H_Y$ in Eq.~\eqref{rap-corr}. In particular, 
we would like to know whether it contains any collinear divergence.  For this purpose, we transform the result to momentum space. 
The term $\propto S^{(2)}(b_\perp, b'_\perp)$ in the bracket $\{\cdots \}$ can be written as follows
\bea 
I_2 = & 2 \int \frac{d^2q_\perp}{q_\perp^2} {\cal F}(x_g, k_\perp)
- 2 \int \frac{d^2q_\perp}{(k_\perp - q_\perp)^2} {\cal F}(x_g, q_\perp)
\nnu
\to &\, 2\pi \int \frac{d^2b_\perp d^2b'_\perp }{(2\pi)^2}e^{-i k_\perp\cdot(b_\perp - b'_\perp)} 
S^{(2)}(b_\perp, b'_\perp)
\nnu
&\times
\left[\frac{1}{\hat\epsilon} + \ln\mu^2 - \ln\frac{c_0^2}{(b_\perp - b'_\perp)^2}\right],
\label{rap-I2}
\eea
where in the second step we use dimensional regularization following 
Eq.~\eqref{DR} with $1/\hat \epsilon=1/\epsilon-\gamma_E+\ln 4\pi$ and $c_0=2e^{-\gamma_E}$. 
On the other hand, the second term $\propto S^{(4)}(b_\perp, x_\perp, b'_\perp)$ in Eq.~\eqref{rap-corr} is given by
\bea
I_4 = & \int d^2\ell_\perp d^2q_\perp\frac{(\ell_\perp-q_\perp)
\cdot (\ell_\perp-k_\perp)}{(\ell_\perp-q_\perp)^2(\ell_\perp-k_\perp)^2} \big({\cal G}(x_g, q_\perp, k_\perp)
\nnu
& + {\cal G}(x_g, k_\perp, q_\perp) \big) - 2 \int d^2\ell_\perp d^2q_\perp
\nnu
& \times \frac{(q_\perp-k_\perp)\cdot (\ell_\perp-k_\perp)}{(q_\perp-k_\perp)^2(\ell_\perp-k_\perp)^2}{\cal G}(x_g, q_\perp,\ell_\perp)
\nnu
\to & \int \frac{d^2b_\perp d^2b'_\perp }{2\pi}e^{-ik_\perp\cdot(b_\perp - b'_\perp)}S^{(2)}(b_\perp, b'_\perp)
\left[\frac{1}{\hat\epsilon}+\ln\mu^2\right]\nnu
&-\pi\int d^2q_\perp \ln(k_\perp-q_\perp)^2 \big({\cal G}(x_g, q_\perp, k_\perp) 
\nnu
&+ {\cal G}(x_g, k_\perp, q_\perp) \big),
\label{rap-I4}
\eea
where ${\cal G}(x_g, k_\perp, q_\perp)$ is defined as
\bea
{\cal G}(x_g, k_\perp, q_\perp)=&\int\frac{d^2b_\perp d^2b'_\perp d^2x_\perp}{(2\pi)^4}e^{-ik_\perp\cdot(b_\perp - x_\perp)}
\nnu
&\times e^{-iq_\perp\cdot(x_\perp - b'_\perp)}S^{(4)}(b_\perp, x_\perp, b'_\perp).
\eea
Finally, we can  write the rapidity factorization correction term in Eq.~\eqref{rap-corr} as
\bea
\Delta H_Y &= \frac{\alpha_sN_c}{\pi} \int_{\tau}^1\frac{dz}{z^2} D_{h/q}(z) x_p f_{q/p}(x_p) \int_{Y_0}^{Y} dy_A
\nnu
&\hspace{-20pt}\times \left[
\int \frac{d^2b_\perp d^2b'_\perp }{(2\pi)^2}e^{-iq_\perp\cdot(b_\perp - b'_\perp)}S^{(2)}(b_\perp, b'_\perp)\ln\frac{c_0^2}{(b_\perp - b'_\perp)^2}
\right.
\nnu
&\hspace{-20pt} - \frac{1}{2} \int d^2q_\perp 
\ln(k_\perp-q_\perp)^2 \big({\cal G}(x_g, q_\perp, k_\perp) + {\cal G}(x_g, k_\perp, q_\perp) \big)
\nnu
& \hspace{-20pt} 
-\left.
\frac{1}{\pi}\int d^2\ell_\perp d^2q_\perp\frac{(q_\perp-k_\perp)\cdot (\ell_\perp-k_\perp)}{(q_\perp-k_\perp)^2(\ell_\perp-k_\perp)^2}{\cal G}(x_g, q_\perp,\ell_\perp)
\right].
\label{finite}
\eea
In other words, the $1/\hat \epsilon+\ln\mu^2$ term cancels between Eqs.~\eqref{rap-I2} and  \eqref{rap-I4}. This indicates that 
the rapidity divergence and collinear divergence are well separated, and thus can be factorized independently.

%%%%%%%%%%%%%%

{\it Numerical results.}
To illustrate  our NLO calculation, we use the  GBW model \cite{GolecBiernat:1998js} to parametrize the dipole 
scattering amplitude: $S^{(2)}(b_\perp, b'_\perp)=\exp\left[-(b_\perp - b'_\perp)^2 Q_s^2(x)/4\right]$.
The saturation scale in a nucleus with atomic number $A$ is given by
$Q_s^2(x) = c A^{1/3} Q_{s0}^2(x_0/x)^{\lambda}$,
with $Q_{s0}=1$ GeV, $x_0=3.04\times 10^{-4}$ and $\lambda=0.288$. We use $c=0.56$ \cite{Stasto:2011ru} for minimum bias p+A collisions. 

\begin{figure}[!t]
\psfig{file=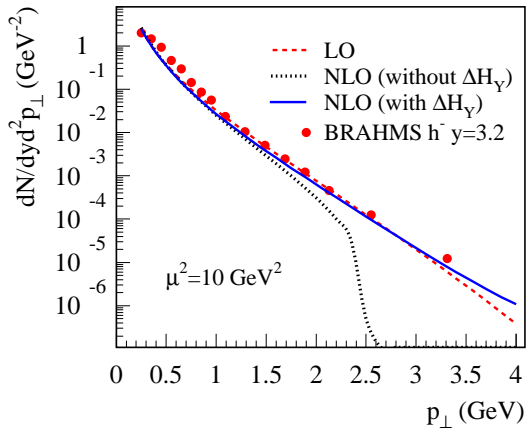, width=2.7in}
\caption{Comparison of $h^-$ spectra obtained in the small-$x$ formalism with fixed $\mu^2 = 10$~GeV$^2$ to BRAHMS data~\cite{Arsene:2004ux}. }
\label{negative}
\end{figure}

We first show that, within our rapidity factorization scheme, the full NLO results with  $\Delta H_Y$ 
in Eq.~\eqref{finite} remedies the negative cross section from  the  calculation in Ref.~\cite{Stasto:2013cha}. 
In Fig.~\ref{negative} we present  comparison to the BRAHMS $h^-$ data at $y=3.2$ in d+Au collisions at 
RHIC~\cite{Arsene:2004ux}. For consistency with~\cite{Stasto:2013cha}, we choose the collinear 
factorization scale $\mu^2=10$ GeV$^2$. The red dashed curve is the LO result, the blue solid curve 
is our NLO calculation (including the new rapidity correction term $\Delta H_Y$), while 
the black dotted curve is the previous NLO result that becomes negative for 
$p_\perp \gtrsim 2.5$~GeV~\cite{Stasto:2013cha}. We have checked that the formalism presented here 
yields positive-definite cross sections for variety of rapidities and center of mass energies
in the  physical kinematic $p_\perp$ region.

\begin{figure}[!t]
\psfig{file=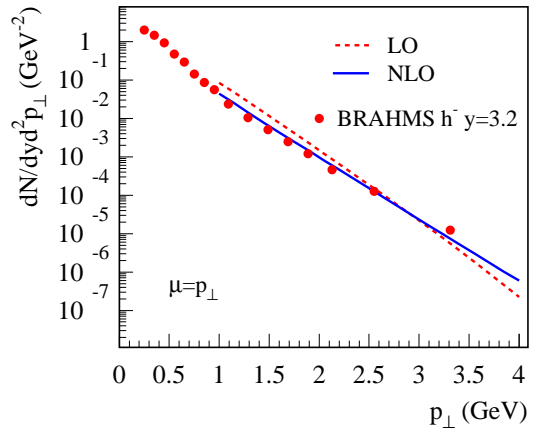, width=2.7in}
\caption{Comparison of the LO and NLO results to BRAHMS data \cite{Arsene:2004ux}. 
We choose the collinear factorization scale~$\mu = p_\perp$.}
\label{mu-pt}
\end{figure}

Of course, one should choose the collinear factorization scale $\mu$ to be related to the 
typical momentum scale in the hard process (e.g.\ $p_\perp$ of the hadron). In Fig.~\ref{mu-pt} 
we plot a new comparison to the BRAHMS data with  $\mu = p_\perp$. The red dashed curve 
shows the LO result, the blue solid curve shows our NLO calculation (with $\Delta H_Y$ included).
At one loop we find a good 
description of the experimental data. At higher $p_\perp$ our NLO corrections enhances the cross section 
as expected, since it includes the gluon radiation processes.

\begin{figure}[!b]
\psfig{file=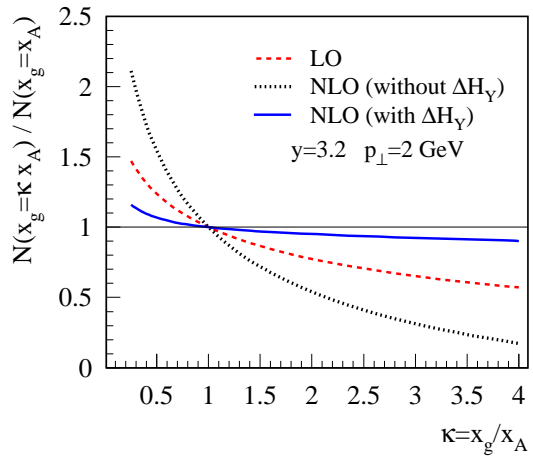, width=2.7in}
\caption{The rapidity factorization scale  $Y_0=\ln1/x_g$ dependence.}
\label{rap-dep}
\end{figure}

As we emphasized earlier, the factorization scale $\mu$-dependence should be largely reduced in 
the NLO cross section when compared to the LO results. We have verified that this is 
indeed the case, consistent with previous findings~\cite{Stasto:2013cha}.
What is much more important is to demonstrate the reduction in sensitivity to the rapidity 
factorization scale $Y_0=\ln1/x_g$. We plot in Fig.~\ref{rap-dep} the ratio 
$R=\left.\left.\frac{dN}{dyd^2p_\perp}\right|_{x_g = \kappa x_A}\right/\left.\frac{dN}{dyd^2p_\perp}\right|_{x_g = x_A}$ as a function of 
$\kappa = x_g/x_A$, with $x_A$ being the typical gluon momentum fraction at LO. 
It can be seen that for $\kappa\in (0.25, 2)$ the LO result has a variation of $\pm 50\%$, while our NLO result 
with the new rapidity correction $\Delta H_Y$ shows only $\pm10\%$ variation. 
On the other hand, the previous result from~\cite{Stasto:2013cha} shows more than a factor of 
2 variation. In other words, the full NLO calculations provide predictions that are much more stable  
with respect to variation of both collinear factorization and rapidity 
factorization scales.

%%%%%%%%%%%%%%
{\it Summary.} In this paper we studied forward hadron production in high energy p+A collisions 
within the small-$x$ formalism. We revisited the previous one-loop calculation and demonstrated that 
besides the well-known collinear factorization, which separates the short-distance from the long-distance physics, 
one has to pay close attention to the ``rapidity factorization'' regime. 
It separates the small-$x$ dynamics of ``fast'' and ``slow'' gluon fields. 
The rapidity factorization procedure results in a new next-to-leading order correction 
which remedies the unphysical negative cross section from the one-loop calculation of~\cite{Stasto:2013cha}. 
We also demonstrated that such factorization formalism leads to much more stable and reliable
cross section predictions at  next-to-leading order. We expect that our results will have 
important applications for  small-$x$ gluon saturation phenomenology. 

This research is supported by the US Department of Energy, Office of Science, and in part by the LDRD program at LANL.

%%%%%%%%%%%%%%


\begin{thebibliography}{99}
\bibitem{scale}
For simplicity, here we have chosen the same factorization scale for both PDFs and FFs. In principle, they don't have to be the same.
\end{thebibliography}
\end{document}